\documentclass[a4paper]{article}

\usepackage[english]{babel}
\usepackage[utf8x]{inputenc}
\usepackage[T1]{fontenc}
\usepackage{array}

\usepackage[a4paper,top=3cm,bottom=2cm,left=3cm,right=3cm,marginparwidth=1.75cm]{geometry}

\usepackage{amsmath}
\usepackage{graphicx}
\usepackage[colorinlistoftodos]{todonotes}
\usepackage{cite}
\usepackage[colorlinks=true, allcolors=blue]{hyperref}
\usepackage{authblk}

\begin{document}

\title{Infrared absorbers inspired by nature}

\date{}

\author[1,2,*]{S\'{e}bastien R. Mouchet}

\affil[1]{Department of Physics, Namur Institute of Structured Matter (NISM) \& Institute of Life, Earth and Environment (ILEE), University of Namur, Rue de Bruxelles 61, 5000 Namur, Belgium}
\affil[2]{School of Physics, University of Exeter, Stocker Road, Exeter EX4 4QL, United Kingdom}
\affil[*]{sebastien.mouchet@unamur.be; s.mouchet@exeter.ac.uk; ORCID: 0000-0001-6611-3794}

\maketitle

\begin{abstract}
Efficient energy harvesting, conversion, and recycling technologies are crucial for addressing the challenges faced by modern societies and the global economy. The potential of harnessing mid-infrared (mid-IR) thermal radiation as a pervasive and readily available energy source has so far no been fully exploited, particularly through bioinspiration. In this article, by reviewing existing photon-based strategies and the efficiency of natural systems in harnessing light and thermal radiation, I highlight the promising role of bioinspiration in enhancing energy capture, conversion, and recycling. Natural photonic structures found in various organisms, including insects, birds, and plants, exhibit sophisticated optical properties that can be leveraged for energy-efficient applications. These developments pave the way for future research and innovation in bioinspired energy solutions. Ultimately, they contribute to the pursuit of a sustainable and environmentally conscious future by harnessing the beauty of nature's designs to meet humankind's energy needs.
\end{abstract}


\section{Introduction}

Addressing the significant challenges faced by modern global society and the world economy, the advancement of efficient energy harvesting and recycling technologies~\cite{Baran2017,Breyer2017,Green2017} stands as a prominent area of research on a global scale. Mid-infrared (mid-IR) thermal radiation represents a pervasive and readily available energy source. This is not only due to the long illumination of some parts of the Earth by the Sun but also because many machinery, engines, and industrial processes dissipate energy in the form of heat radiation, distinct from thermal conduction or convection mechanisms.

While the primary energy source may vary in its environmental impact, the recycling of this "wasted" energy presents a sustainable approach to converting radiative heat losses into diverse forms of energy. Numerous mechanical components found in machinery, engines, industrial processes, and even household systems generate mid-IR thermal emissions at moderately elevated temperatures, typically ranging from 150°C to 950°C. These emissions are an intrinsic byproduct of these components' regular functioning and constitute an unavoidable energy loss. The prospect of harnessing this radiative heat loss is compelling, as it offers the opportunity to transform it into electrical power, effectively enabling devices to utilise their own recycled radiative heat loss for enhanced functionality.

Photon-based strategies have already played a crucial role in harnessing solar energy, enhancing the performance of energy conversion devices~\cite{Yablonovitch1982, Niggemann2008,Herman2012,Brongersma2014,Mayer2014,Deparis2015,Mayer2015,Rahman2015,Madanu2023b}. For instance, devices designed for solar light trapping have effectively increased the efficiency of photovoltaic (PV) cells and thermal photovoltaic (TPV) cells. Similar photonic devices are instrumental in augmenting the efficiency of solar thermal panels, or in energy harvesting for thermoelectric generators (TEG), artificial photosynthesis, and photocatalysis.

In nature, numerous biological organisms have developed highly efficient mechanisms to harness thermal radiation, a crucial adaptation for their survival. Over millions of years of evolution, these natural systems have honed specialised characteristics to maximise their radiation harvesting abilities~\cite{mouchet2018structural,Mouchet2021}. Consequently, certain structures within their integuments have become increasingly inspiring for the development, design, and production of energy-efficient materials~\cite{Biro2011,Deparis2014,Zhou2017,Mouchet2021}. Bioinspiration emerges as a powerful and promising strategy in this context.

Natural photonic structures found in various animals, including insects, birds, and fish, are example of effective thermal radiation collectors~\cite{Biro2011,Zhou2017,Yoshida2002}. In addition, this type of structures exhibit a diverse array of properties, such as structural colours (resulting from light interference in nanostructures)~\cite{Mouchet2012b,Pasteels2016,mouchet2018structural,Mouchet2021}, antireflection features~\cite{Yoshida2002,Stavenga2006,Deparis2014}, thermoregulation mechanisms~\cite{Biro2003,Berthier2005,StuartFox2017,Liu2019,Xie2019}, light-trapping capabilities~\cite{Herman2011,Vukusic2004,Han2015,Tian2015}, and enhanced light-extraction methods~\cite{Bay2013}. These properties emerge from the interaction between radiation and structures composed of biopolymers like chitin, keratin, collagen, or cellulose, sometimes in combination with pores.

The existence of these naturally occurring radiation management systems challenges human imagination. While human beings have access to a wide range of materials, human designs sometimes fall short in complexity compared to these remarkable natural structures. Identifying and comprehending these natural photonic devices not only expand human understanding but also empowers engineers and materials scientists to conceptualise new ideas and explore potential technological applications through bioinspired principles~\cite{Biro2011,Deparis2014,Zhou2017,Mouchet2021}. These exciting possibilities have captivated the attention of researchers worldwide. Despite the development of artificial intelligence, bioinspiration remains a guiding force in the quest for novel technological applications. In fact, the convergence of both approaches holds promise for unprecedented advancements in this field.

In this article, I first review previously investigated cases of photonic structures enhancing electromagnetic-wave absorption (also known as structural absorption) in natural organisms across the UV, visible and infrared (IR) range. This is because the dimensions of a visible light absorber occurring in nature may be adjusted to another range such as IR through a bioinspiration approach due to the scalability of Maxwell's equations. Finally, I review examples of bioinspired IR absorbers from the literature.

\section{Light absorption enhanced by photonic structures in natural organisms}

The management of electromagnetic radiation and thermoregulation are pivotal functions essential for the survival or benefice of various natural organisms, including plants, insects, and birds~\cite{Berthier2005,NanShi2015,Shanks2015,Hunig2016,Jacobs2016,StuartFox2017,McCoy2018,Liu2019,Xie2019,Xu2024}, whether endotherms (organisms able to maintain their body temperature through their metabolisms), mesotherms (organisms with some metabolic strategies of heat production without any proper metabolic heat control), or ectotherms (organisms requiring external heat sources)~\cite{Cossins2012}. For instance, photosynthesis implies absorbing visible radiation from the Sun whereas thermoregulation of ectothermic animals involves a subtle trade-off between radiation absorption and thermal emission in the near-infrared (near-IR) part of the electromagnetic spectrum. Photonic structures may play roles in the management of such thermal radiation. For instance, iridescent butterflies were reported to exhibit in general an absorptance higher than the one of non-iridescent species~\cite{Bosi2008}. Other striking illustrations are the photonic structures occurring in the super-black feathers of the bird of paradise (as depicted in figure~\ref{fig:BlackBird})~\cite{McCoy2018}, as well as in the scales covering of the black wings of insects like the Magellan birdwing and the Meander prepona butterflies~\cite{Berthier2005,Herman2011}. These feathers and wings exhibit remarkably high energy absorption properties within the spectral range of solar irradiance, encompassing the mid-IR spectrum in some instances. Often, in such natural integument, incident light is absorbed by pigments including melanin~\cite{Slominski2004,Gunderson2008,NilssonSkold2016,Mouchet2023b}. Nanostructures, operating at micro- and nanometer length-scales, yield fascinating opportunities for both light and thermal radiation harvesting.

\subsection{Wings of lepidopterans} \label{winglepidopterans}

\begin{figure}
    \centering
    \includegraphics[width=135mm ]{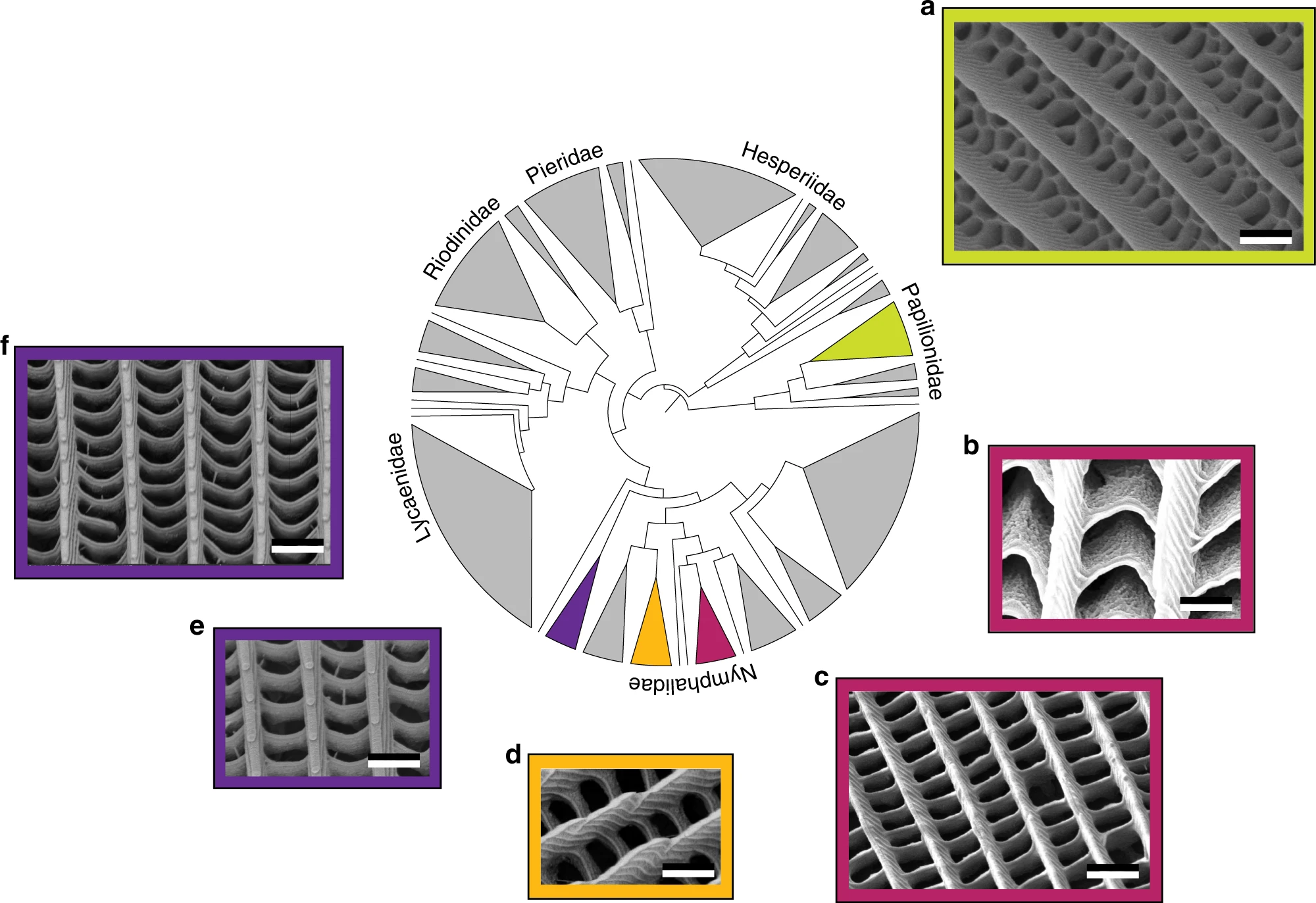}
    \caption{Diverse structures of scales exhibit ultra-black colouration within the order Lepidopetra. They typically comprise holes located in between the scales' ridges with various sizes and shapes: honeycomb (a), chevrons (b), and rectangles (c-f), as observed by SEM with wings of \textit{Trogonoptera brookiana} male papilionid (a), \textit{Eunica chlorocroa} nymphalid (b), \textit{Catonephele antinoe} nymphalid (c), \textit{Heliconius doris} nymphalid (d), \textit{Euploea dufresne} nymphalid (e), and \textit{Euploea klugi} nymphalid (f). Scale bars: 1~$\mu$m (a-f). This figure was reproduced from ref.~\cite{Davis2020}, License CC-BY-4.0.}
    \label{fig:papiliophylogen}
\end{figure}

Insects such as lepidopterans, the taxonomic order encompassing the ethereal beauty of butterflies and the enigmatic allure of moths, are ectotherms, which means that their metabolisms rely on environmental heat sources. Harvesting incident energy appears crucial for these species. The phylogenetic diversity of structures giving rise to ultra-black colouration occurring in the order Lepidopetra was recently analysed in detail (figure~\ref{fig:papiliophylogen})~\cite{Davis2020}. All these structures present some longitudinal ridges connected by crossribs in the upper lamina of the scales, forming 2D networks of quasi-periodic holes. The resulting high-surface area was described as increasing light absorption by underlying cuticular melanin and reducing reflection~\cite{Yan2016,Davis2020}: whatever the size and shape of the holes -honeycomb, chevrons, or rectangles- scales giving rise to ultra-black visual appearances exhibit steeper ridges as well as deeper and wider trabeculae (namely, pillars connecting the upper and basal laminae of a scale) than scales with some regular black or brown colour. Through numerical modelling, these features were shown to play a significant role in reducing light reflection~\cite{Yan2016,Davis2020}. Furthermore, coating these structures with gold does not lead to an increase in light reflectance, unlike regular black or brown butterfly wings. This experimentally demonstrates the photonic origin of the related light harvesting~\cite{Davis2020}.

\begin{figure}
    \centering
    \includegraphics[width=135mm]{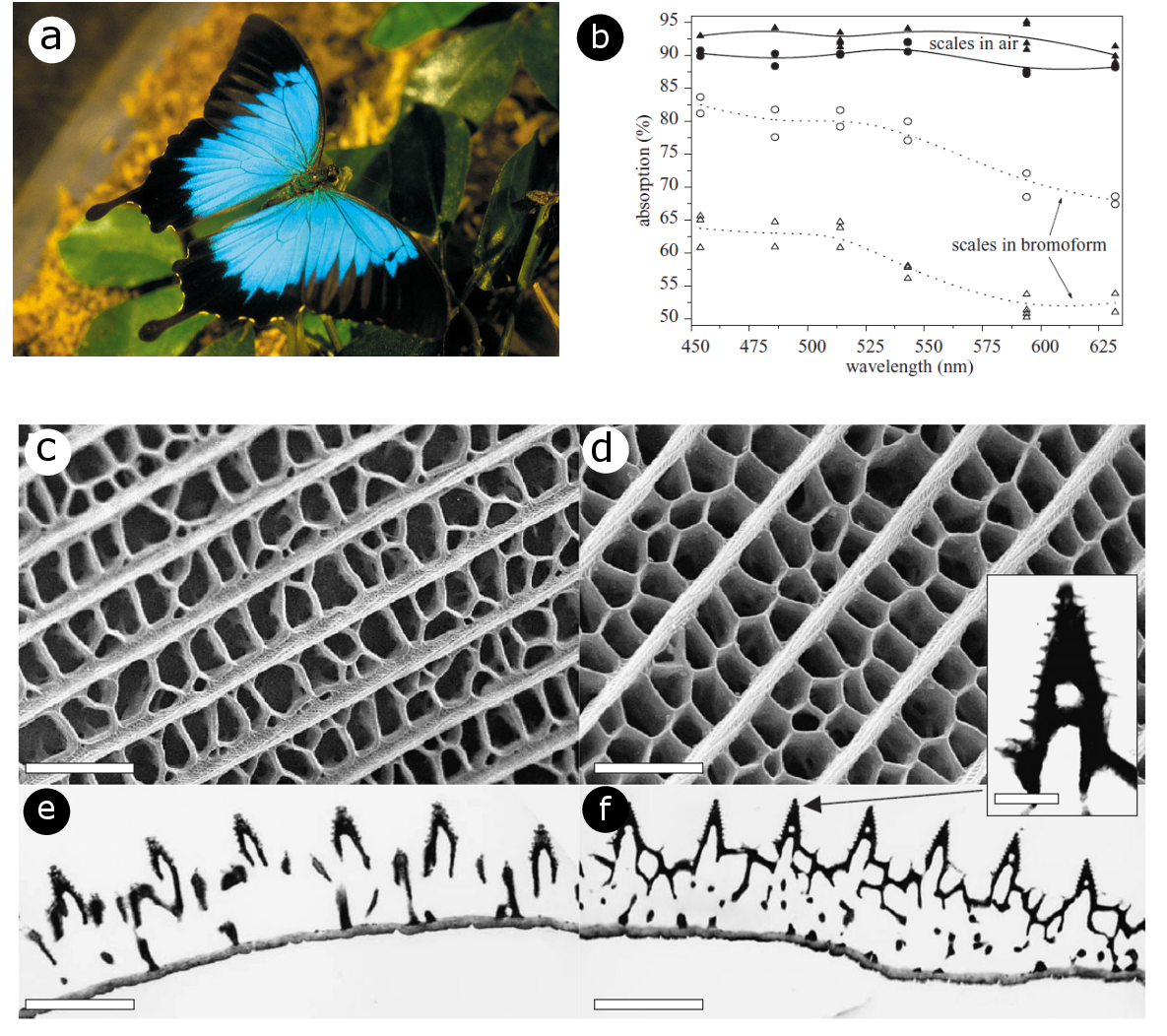}
    \caption{The male \textit{P. ulysses} butterfly (a) exhibits some black areas on its wings. Upon contact with index-matching fluid (here, bromoform), the absorption spectra exhibit significantly lower intensities (b). They were measured at normal incidence with both lustrous (circles) and matt (triangles) black scales. Electron microscopy (c,d: SEM; e,f: TEM) allowed to observed the structures of the lustrous (c,e) and matt (d,f) black scales. Scale bars: 3~$\mu$m (c), 2~$\mu$m (d); 2~$\mu$m (e,f); inset, 300~nm. These figures were reproduced from W. van Aken (a), https://commons.wikimedia.org/wiki/File:CSIRO\_ScienceImage\_3831\_Ulysses\_Butterfly.jpg, License CC-BY-3.0, and from ref.~\cite{Vukusic2004} (b–f), with permission from the Royal Society.}
    \label{fig:papilioulysses}
\end{figure}

In the scales covering the wings of the male \textit{Papilio ulysses} butterfly, similar complex photonic structures were found to play a role in intensifying the darkness areas~\cite{Vukusic2004}, in addition to the porous multilayer structure producing the bright blue colour through interference~\cite{Vukusic2001}. They were described to scatter light towards the ridges and the interior of the scale, leading to a longer optical path, resulting in a higher light absorption by the pigments distributed within the scale material. For matt black scales, the absorption reduced from 95\% to 55\%, upon contact with bromoform serving as index-matching fluid, while for lustrous black scales, it decreased from 90\% to 70\% (figure~\ref{fig:papilioulysses}). Ridges were however demonstrated through simulation to have neglected role in visible-light absorption of the ultra-black scales of \textit{Pachliopta aristolochiae}~\cite{Siddique2017}.

The wings of \textit{Troides magellanus} butterfly, the Magellan birdwing, present a captivating spectacle of optical properties including light diffraction and controlled fluorescence emission on their hindwings~\cite{Vigneron2008,Lee2009,VanHooijdonk2012,VanHooijdonk2012b}. It inhabits the Philippines and Taiwan's Orchid Island. Renowned for their impressive size and striking appearance, the forewings showcase a remarkable 98\% absorption of visible light as well as reveal two distinctive peaks in the IR spectrum~\cite{Herman2011}. The presence of chitin imparts the wings with these robust absorption peaks at 3~$\mu$m and 6~$\mu$m due to C = O vibrations, strategically positioned within the wavelength range where a black body emits radiation at 40°C, enabling radiative cooling. The architecture of the Magellan birdwing consists of five major elements~\cite{Herman2011}: a roof-like structure on which are located a series of ridges, holes in the separating structures and pillars, joining the upper membrane to the lower membrane of the wing. A similar structure was also found in the case of the related \textit{Troides aeacus}~\cite{Zhao2011}. Comparison of numerical simulations between the photonic structure and a non-structured flat slab with equal volume of material showed a 10\% increase in electromagnetic radiation absorption and a 17\% increase in emissivity at 40°C~\cite{Herman2011}. This unique combination of optical characteristics suggests that the Magellan birdwing has evolved to manage efficiently both visible and IR light, underscoring the sophisticated adaptation of these butterfly wings for light and thermal radiation purposes.

\begin{figure}
    \centering
    \includegraphics[width=135mm ]{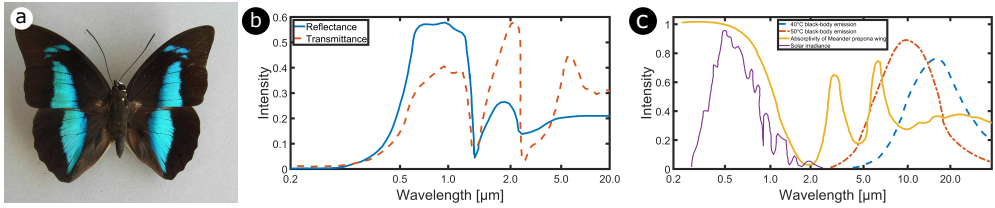}
    \caption{Radiative cooling plays a key role in the thermoregulation of certain butterfly wings, exemplified by \textit{Archaeoprepona meander}, the Meander prepona. It arises from the morphology of the scales covering the black area of the wings, as observed here by SEM. A clear difference in reflectance and transmittance was measured between the near-IR and visible parts of the electromagnetic spectrum (c). Black body emission spectra at 40°C and 50°C, solar irradiance, and the absorptivity of a wing (d) allow to explain the thermoregulation enabled by the C = O absorption peak of chitin at 6~$\mu$m. Figure (a) was reproduced from Notafly, https://commons.wikimedia.org/wiki/File:Archaeoprepona\_meander\_(Cramer,\_-1775-).JPG, License CC-BY-SA-3.0. Data in figures (b,c) are from ref.~\cite{Berthier2005}.}
    \label{fig:thermoregulation}
\end{figure}

A related thermoregulation was demonstrated in the case of \textit{Archaeoprepona meander}, the Meander prepona, a tropical butterfly species (figure~\ref{fig:thermoregulation})~\cite{Berthier2005}, as well as later on in various butterfly species~\cite{Krishna2020}. These insects employ a sophisticated mechanism to manage its body temperature effectively within a given range such as 36°C-40°C~\cite{Berthier2005} or 20°C-50°C~\cite{Krishna2020}, depending on the species. The intricate structure of their wings such as the black wings of Meander prepona serves as a remarkable example of nature's engineering prowess. The scale structures on the wings, in addition to melanin pigments, play a crucial role in harnessing solar energy efficiently, absorbing approximately 95\% of the visible solar spectrum (figure~\ref{fig:thermoregulation}c)~\cite{Berthier2005}, akin the phenomenon described in the cases of \textit{Papilio ulysses} and the Magellan birdwing hereabove. In the near-IR range, the absorptance intensity decreases down to less than 2\%, ensuring low thermal emissivity, apart from the absorptance peaks at 3~$\mu$m and 6~$\mu$m. The 6-$\mu$m emissivity peak plays a crucial role in thermoregulation~\cite{Berthier2005}. At temperatures below 40°C, the black body peak is located at a longer wavelength (figure~\ref{fig:thermoregulation}c). It allows the wing to harvest heat effectively while maintaining low thermal emission. When temperatures exceed 40°C, a significant overlap between the black body spectrum and the 6-$\mu$m peak appears (figure~\ref{fig:thermoregulation}c), leading to higher thermal emission (namely, radiative cooling) and contributing to the butterfly's fine-tuned response to thermal challenges in its habitat. This thermoregulation mechanism could be employed in applications such as solar energy harvesting as it can help maintaining the devices within an optimal temperature range.

The role of ultra-black colours in butterflies remains the subject of speculation. However, it was hypothesised that they enhance the contrast in the visual signals, as ultra-black areas are always located next to bright areas~\cite{Davis2020}. Such visual contrast would have implications in terms of aposematism or intraspecific communication.

\begin{figure}
    \centering
    \includegraphics[width=135mm ]{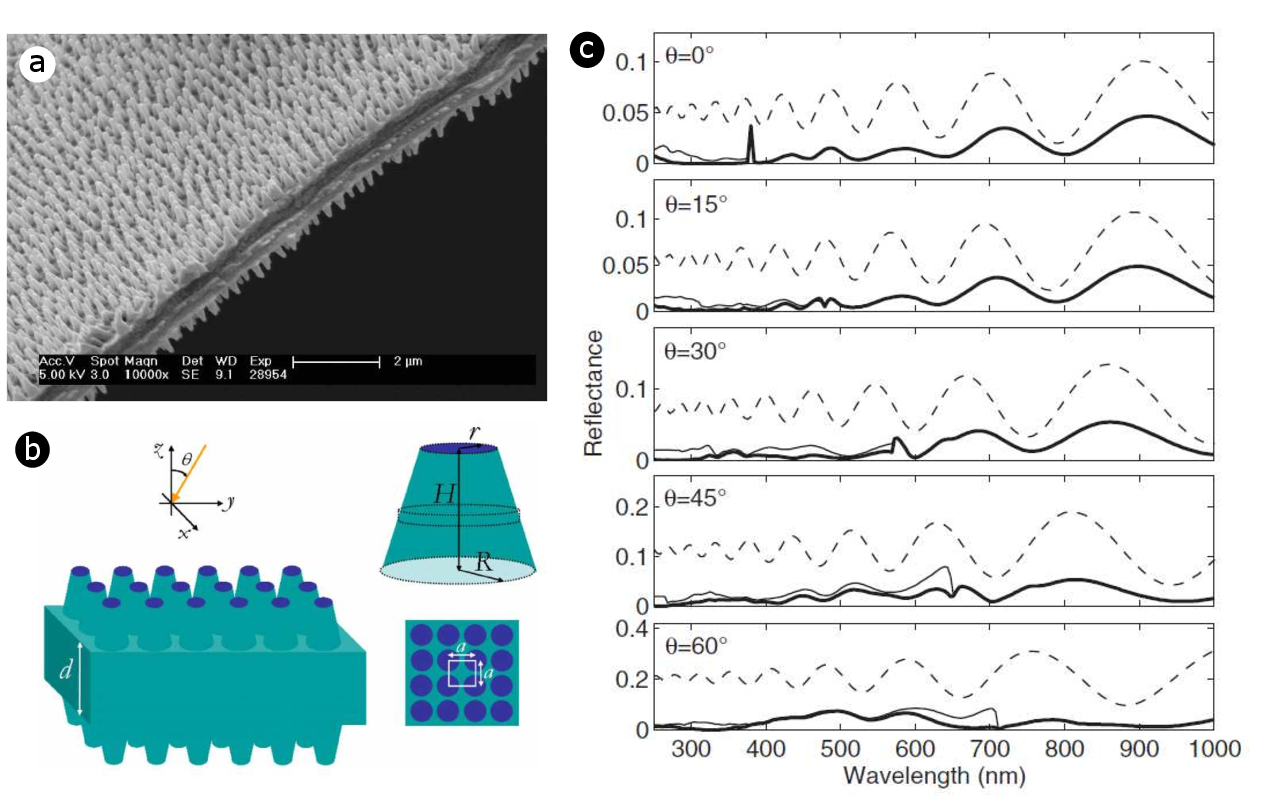}
    \caption{\textit{Cacostatia ossa} moth's wings exhibit increased transparency due to a unique disordered structure in nipple array covering them (a). This structure was modelled by truncated cones for reflectance simulations (b). Its total (thin solid curves) and specular (thick solid cuves) reflectance spectra (c) at different angles of incidence are much lower than the specular reflectance of a unstructured flat wing (dashed curves). These figures were reproduced from ref.~\cite{Deparis2009} with permission from the American Physical Society.}
    \label{fig:Deparis_nipple_array}
\end{figure}

In addition to structured scales covering the ventral and dorsal sides of lepidopteran wings, many species including butterflies \textit{Greta} spp., the moth \textit{Cacostatia ossa}, and the moth \textit{Cephonodes hylas} display some highly transparent scale-less wings with antireflection properties through photonic structuring of the wing membranes~\cite{Yoshida1996,Yoshida1997,Yoshida2002,Vukusic2003,Deparis2009,Deparis2014,Stavenga2014,Siddique2015,Mouchet2023}. This structuring curtails reflection of incident light to levels below 2\% across the entire visible spectrum, through electromagnetic impedance matching. It consists in a lattice of dome-shaped protuberances, also known as nipples (figure~\ref{fig:Deparis_nipple_array}). The underlying principle behind this antireflection effect lies in the gradual refractive index matching between the air and the wing membrane, typically composed of chitin. If the protuberances are spaced by less than the incident wavelength, typically less than 200~nm, the non-zero diffraction orders are evanescent. The protuberance structure can be regarded as a slow variation of the effective refractive index along the normal to the wing membrane. Depending on the species, the protuberance lattice can be very well ordered such as the hexagonal compact array in the wings of \textit{C. hylas}~\cite{Yoshida1996,Yoshida1997,Yoshida2002} and \textit{Hemaris fuciformis}~\cite{Vukusic2003,Mouchet2023} hawkmoths or more disordered such as the wings of \textit{C. ossa} moth (figure~\ref{fig:Deparis_nipple_array})~\cite{Deparis2009,Deparis2014} and the ones of \textit{Greta} spp. glasswing butterflies~\cite{Stavenga2014,Siddique2015}. Interestingly, disorder in the protuberance height, width, and position was found to increase the transparency properties, in the case of \textit{G. oto} glasswing butterfly~\cite{Siddique2015}. Beyond the order of lepidopterans, antireflection structures manifest in the wings of odonatans, such as \textit{Aeshna cyanea} dragonfly~\cite{Hooper2006,Mouchet2023}, the American Rubyspot dameselfly \textit{Hetaerina americana}~\cite{Stavenga2014}, and \textit{Vestalis amabilis} damselfly~\cite{Mouchet2023}, or even in hemipterans like cicadas~\cite{Stoddart2006,Sun2011,Dellieu2014,Deparis2014,Verstraete2019,Mouchet2023}. In addition, such nipple arrays were observed on the surfaces of compound-eye corneas of several arthropods~\cite{Bernhard1965,Bernhard1970,Parker1998,Vukusic2003,Stavenga2006}. They are often referred to as moth-eye structures. A comparative study of 19 species of butterflies led to the classification of the arrays into three categories according to their morphologies: conical, paraboloidal, and Gaussian. The paraboloid profile with protuberances almost touching each other was found to exhibit the lowest reflectance, with the effective refractive index varying quasi-linearly with depth~\cite{Stavenga2006}. Highly antireflective wings in insect are often reported to play a likely role in crypsis~\cite{Sun2011,Dellieu2014}. Similarly, nipple arrays on the surfaces of compound eyes are assumed to improve camouflage under daylight and improve night vision~\cite{Bernhard1965,Bernhard1970}. In general, some of such lattices of protuberances were reported to combine antireflection with hydrophobic properties~\cite{Sun2011,Dellieu2014,Deparis2014,Zhang2006,Sun2009,Wisdom2013,RomanKustas2020}, bactericidal activity~\cite{Ivanova2012,Ivanova2013,Diu2014,Kelleher2016,RomanKustas2020}, and fluorescence emission~\cite{Gorb1999,Appel2011,Appel2015,Chuang2016,Mouchet2023}. In the case of cicadas, it was shown that the protrusions could be approximated by truncated cones under hemispheres~\cite{Dellieu2014,Deparis2014}. The cones gave rise to impedance matching and high antireflection whereas the cones favoured hydrophobicity. Fluorescence arises from fluorescent proteins -typically resilin- embedded withing the membrane material~\cite{Gorb1999,Appel2011,Appel2015,Chuang2016,Mouchet2023}.

\subsection{Elytra of beetles}

\begin{figure}
    \centering
    \includegraphics[width=135mm ]{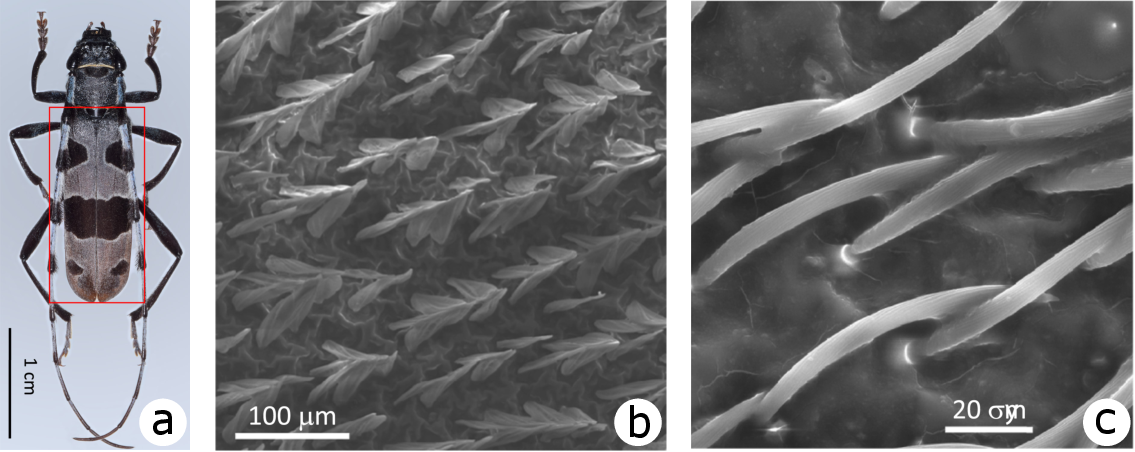}
    \caption{The elytra of \textit{Rosalia alpina} longhorn beetle exhibit black patches on a blue-grey background (a). The black patches are covered by so-called "tent-shaped" scales (b), whereas hairs cover the blue-grey area (c). These figures were reproduced from ref.~\cite{Pavlovic2018} with permission from Elsevier.}
    \label{fig:beetle}
\end{figure}

The blue-grey elytra of \textit{Rosalia alpina} longhorn beetle (family Cerambycidae) exhibit large black spots (figure~\ref{fig:beetle}). The micro- and nanostructured setae that cover these elytra contribute, on one side, to the camouflage of this beetle on beech barks and, on the other side, to thermoregulation by allowing a quick heating of the body to the optimal temperature and dissipating excess of heat through IR emission to prevent overheating~\cite{Dikic2016,Pavlovic2018}, akin the wings of some butterflies described in the previous section. The setae occurring in the black spots enhance visible light absorption by light trapping, whereas the setae of all the elytra enable the thermoregulation. The former are inclined scales, touching neighbours at the tips and forming tent-like architectures with 1-$\mu$m period and 100-nm period grating patterns (figure~\ref{fig:beetle}a)~\cite{Dikic2016,Pavlovic2018}. The setae occurring on the blue-grey areas consist in hairs~\cite{Dikic2016,Pavlovic2018} (figure~\ref{fig:beetle}b). Through optical modelling based on SEM observations, the light trapping role of the scales was demonstrated~\cite{Dikic2016,Pavlovic2018}. Several reflections on opposite inclined patterned scales and high concentration of melanin in these scales account for the high-absorption properties. In addition, the scales and the hairs exhibit absorption (and hence emission) enhancement in the mid-IR range~\cite{Dikic2016,Pavlovic2018}.

More recently, Vasiljević and co-workers unveiled a combination of lenslets and micron-sized multilayered spherical black elements located within the elytra of \textit{Morimus asper funereus} longhorn beetle (family Cerambycidae)~\cite{Vasiljevic2021}, which also display black spots on a grey surface. However, in this case, both areas, black and grey, look identical when observed with a thermal camera. The authors concluded from Finite Element Method (FEM) simulations that the combined action of the lenslets and the multilayered spherical elements focuses IR radiation on microchannels containing hemolymph. 

\begin{figure}
    \centering
    \includegraphics[width=135mm]{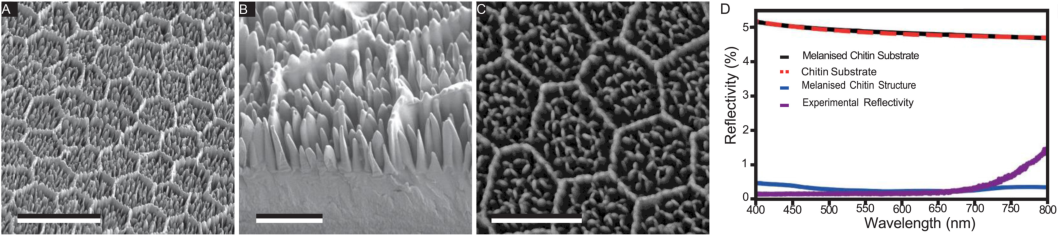}
    \caption{The microstructure occurring on the elytra of \textit{Euprotaetia inexpectata} scarab beetle consists in some ellipsoidal and apparently randomly positioned micropillars (A-C) observed here by SEM with different viewing angles. Numerical predictions of the spectral reflectance from a flat absorbing chitin substrate (black solid line), a flat non-absorbing chitin substrate (red dashed line), and the described absorbing microstructure (blue solid line) confirm the role of the observed structure in the absorption enhancement. The latter simulations agree with experimental results (purple solid line). Scale bars: 15~$\mu$m (A), 5~$\mu$m (B), and 10~$\mu$m (C). These figures were reproduced from ref.~\cite{Parisotto2023}, License CC-BY-4.0.}
    \label{fig:beetle_bodo}
\end{figure}

Finally, arrays of ellipsoidal and randomly located micropillars (figure~\ref{fig:beetle_bodo}A-C) were reported on the elytra of \textit{Euprotaetia inexpectata} scarab beetle (family Scarabaeidae)~\cite{Parisotto2023}. They enhance light absorption by a combination of Mie scattering and optical focusing. This way, incident light reaches absorbing pigment -namely melanin- located within the elytra, giving rise to an absorptance up to 99.5\% and a reflectance of 0.1\% at 400~nm (figure~\ref{fig:beetle_bodo}D).

\subsection{Bird feathers}

\begin{figure}
    \centering
    \includegraphics[width=135mm ]{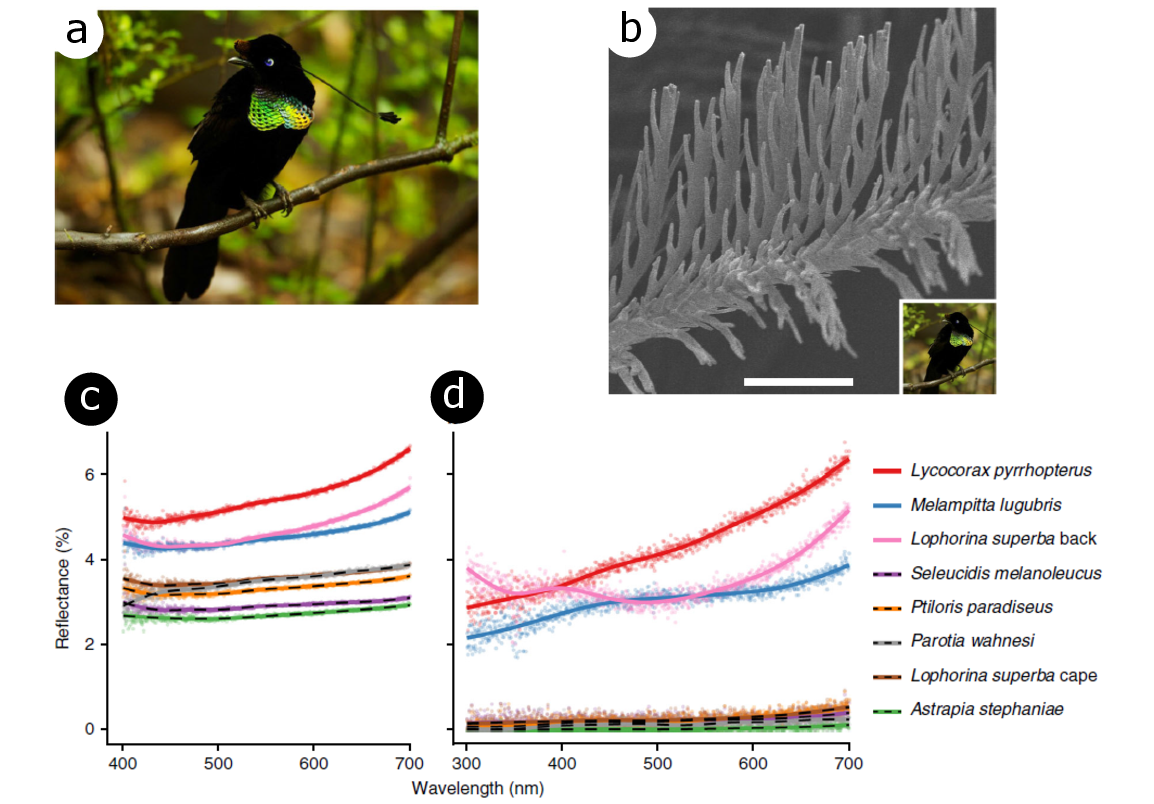}
    \caption{The ultra-black feathers of the \textit{Parotia wahnesi} bird of paradise (a) is characterised by specialised barbules arrays, as observed by SEM (b). Reflectance spectra (c,d) of standard black (solid lines) and ultra-black (dotted lines) feathers are compared, showcasing total (namely, sum of diffuse and specular components) reflectance (c) as well as specular reflectance at normal incidence (d). Scale bar:  50~$\mu$m (b). These figures were reproduced from ref.~\cite{McCoy2018}, License CC-BY-4.0.}
    \label{fig:BlackBird}
\end{figure}

The ultra-black plumage occurring in some species of birds of paradise within the family Paradisaeidae has captivated researchers due to its unparalleled darkness, reaching absorption levels of up to 99.95\%. This phenomenon, elucidated by McCoy and colleagues, is a result of structural absorption rather than pigmentation (figure~\ref{fig:BlackBird})~\cite{McCoy2018}. These feathers appear even darker than typical black feathers due to a significant reduction in specular reflection, as measured through directional reflectance ranging from a mere 0.05\% to 0.31\%. The secret lies in the microstructure of the feathers, featuring barbules curved up that are tilted vertically by ca. 30° with respect to the normal in the direction of the feathers' distal tip. This unique arrangement enhances multiple light scattering, creating regularly spaced cavities with dimensions of 5–30~$\mu$m in width and 200–400~$\mu$m in depth. Astonishingly, the super-black effect is most pronounced when looked from the distal direction, aligning perfectly with the perspective of a female observing a male. The cavities present a directional reflectance bias, making the feathers even darker when viewed from the distal direction. This natural adaptation showcases the fascinating ways in which birds of paradise have evolved to achieve remarkable visual effects in their plumage.

\subsection{Cuticle of \textit{Maratus} jumping spiders}

\begin{figure}
    \centering
    \includegraphics[width=135mm ]{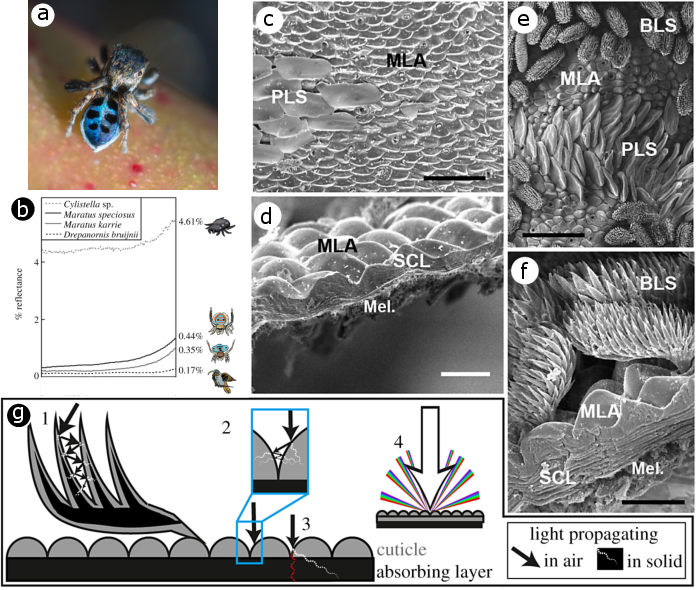}
    \caption{Some male jumping spiders such as \textit{Maratus nigromaculatus} exhibit ultra-black areas along with striking colour for courtship (a). These ultra-black areas reflect 0.44\% and 0.35\% of light with a 30°-detection angle (b) in the cases of \textit{Maratus speciosus} and \textit{Maratus karrie}, respectively. This is significantly less than light reflection from the cuticle of a standard black spider such as \textit{Cylistella} sp. (namely, 4.61\%) and more than the one from the black feathers of \textit{Drepanornis bruijnii} bird of paradise (i.e., 0.17\%). The ultra-blackness of spiders such as \textit{M. speciosus} arises from microlens arrays (MLA) covering some striated layers in the cuticle (SLC) and some absorbing layer of melanin (Mel.) observed in (c,d). Some plate-like scales (PLS) giving rise to to blue colour are also imaged in (c). In the case of \textit{M. karrie} (e,f), some additional brush-like scales (BLS) occur on the surface of the integument. They enhance further light absorption. McCoy and co-workers described four mechanisms of light absorption: 1) Light is scattered multiple times as it interacts with spiny projections of the BLS, gradually being absorbed as it traverses through the cuticle material and into the absorbing melanin layer while scattering; 2) Multiple scatterings occur between bump surfaces, causing light to absorb as it passes through the cuticle materials and enters the melanin layer; 3) The path length of light within melanin layers is extended, leading to increased absorption; 4) Light undergoes diffraction due to a periodic microlens array, resulting in reduced visual detection by an observer such as a female spider. Scale bars: 30~$\mu$m (c), 10~$\mu$m (d), 50~$\mu$m (e), and 10~$\mu$m (f). These figures were reproduced from Graham Wise (a), https://commons.wikimedia.org/wiki/File:Maratus\_nigromaculatus\_(14585680722).jpg, License CC-BY-2.0 and from ref.~\cite{McCoy2019} (b-g), License CC-BY-4.0.}
    \label{fig:JumpingSpider}
\end{figure}

Jumping spiders, specifically the male members of the genus \textit{Maratus}, commonly referred to as peacock spiders, have evolved a fascinating display strategy to attract their female counterparts (figure~\ref{fig:JumpingSpider}a). These spiders exhibit a striking combination of brilliant colours arising from pigments or photonic structures~\cite{Stavenga2016} and velvety black areas~\cite{McCoy2019} on their bodies. These black regions, described as ultra-black~\cite{McCoy2019}, reflect less than 0.5\% of light, reaching intensities as low as 0.35\% in the case of \textit{Maratus karrie}, due to microstructures, including densely packed cuticular bumps resembling microlens arrays (figure~\ref{fig:JumpingSpider}c,d). In addition, \textit{M. karrie} displays some black scales reassembling brushes (figure~\ref{fig:JumpingSpider}e,f). Optical modelling revealed a delicate balance between minimising light reflection from the surface and maximising absorption by melanin (figure~\ref{fig:JumpingSpider}g). Interestingly, McCoy and co-workers proposed that this ultra-black followed a convergent evolution for the success of these spiders and the birds of paradise in the competitive realm of sexual selection~\cite{McCoy2019}.

\subsection{Skin of West African Gaboon viper}

\begin{figure}
    \centering
    \includegraphics[width=135mm ]{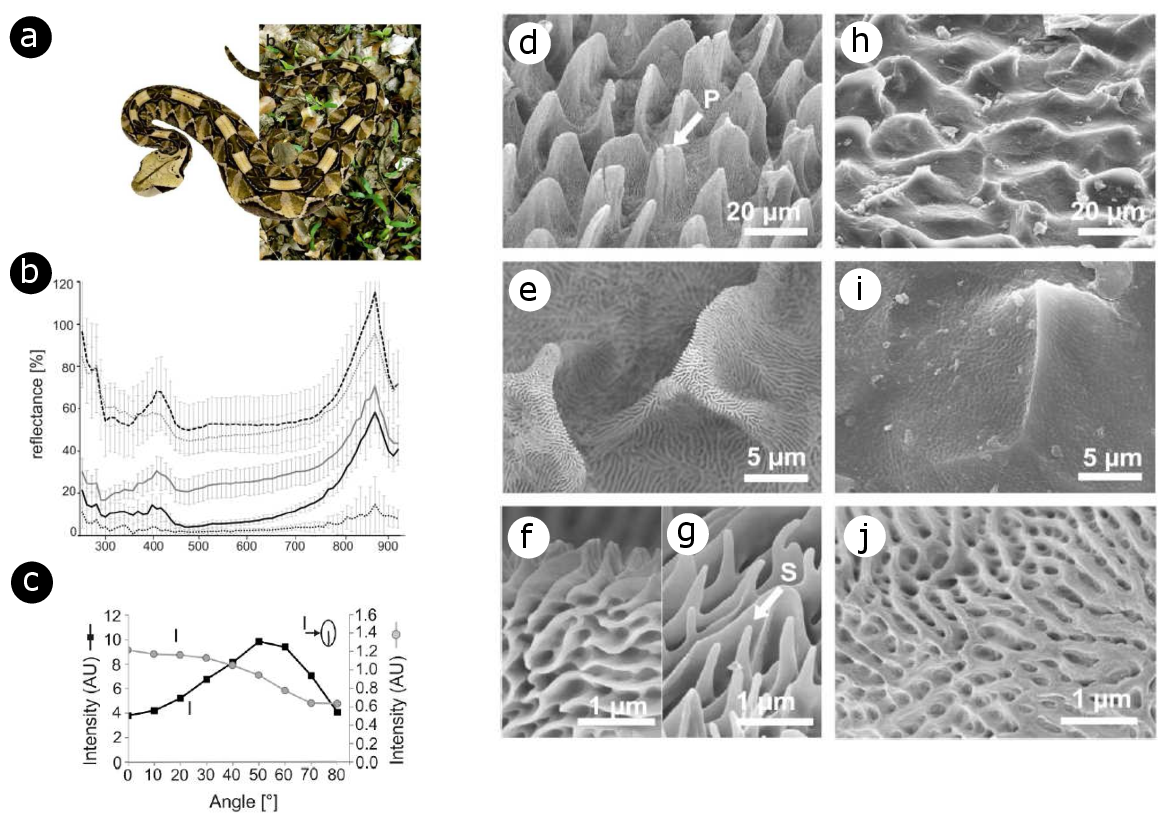}
    \caption{The West African Gaboon viper, \textit{Bitis rhinoceros}, (a) exhibits an effective camouflage pattern with velvety black, light brown and pale hue areas on its skin. Reflectance spectra (b) highlight the characteristics of black dorsal scales (solid black line), Au-Pd coated black dorsal scales (dotted black line), pale dorsal scales (solid grey line), Au-Pd coated pale dorsal scales (dotted grey line), and ventral scales (dashed black line). Unlike pale scales (solid black line), black scales (solid grey line) do not specularly reflect light (c): with an 700~nm incident light at 45° with respect to the normal to the skin surface, light reflectance decreases with the detection angle (c). The microstructures of the scales can be imaged by SEM (d-j). The black scales are densely packed and resemble leaves marked with an arrow (P) in (d). They are covered with small ridges (e-g) with spinules indicated with an arrow (S) in (g). The pale scales have a more simple pattern (h), exhibiting pits (i,j). These figures were reproduced from ref.~\cite{Spinner2013}, with permission from Springer Nature.}
    \label{fig:Viper}
\end{figure}

The Gaboon viper \textit{Bitis rhinoceros}, native to West Africa, exhibits a stunning camouflage in its natural habitat, thanks to its intricate skin pattern~\cite{Spinner2013}. The geometrically arranged velvet black spots, interspersed with pale and light brown regions (figure~\ref{fig:Viper}), seamlessly blend into the diverse light and shade patterns of the forest ground under the canopy. Observations revealed that the blackness of the viper's scales is primarily derived from a hierarchical structure characterised by densely packed, leaf-like microstructures covered with nanoridges. Under microscopic scrutiny, even the areas in between black scales exhibit nanoridge striations (figure~\ref{fig:Viper}). Reflectance spectra analysis demonstrates that both black and pale scales have nearly flat profiles across the visible range, with a notable peak around 880~nm (figure~\ref{fig:Viper}). Intriguingly, applying an Au-Pd coating to the black scales preserves the black colour and further diminishes reflectance (figure~\ref{fig:Viper}). This finding supports the idea that the viper's original surface works as an effective light-trapping device, utilising multiple reflections of light. The metallic coating further enhances light trapping via light reflections on the metal-coated surfaces. Modelling of diffuse reflection using Lambertian symmetric V-shaped cavities validated the proposed light-trapping mechanism and elucidated the angular dependence of reflectance spectra in pale scales~\cite{Spinner2013}. However, the black scales exhibit a distinct angular characteristic, lacking a specular reflection peak and displaying a gradual decrease in reflectance intensity with increasing emerging angle. This unique angular behaviour imparts a nonglossy visual appearance to the velvet black, attributed to the more isotropic arrangement of scale structure.

\subsection{Plant leaves and petals}

\begin{figure}
    \centering
    \includegraphics[width=135mm ]{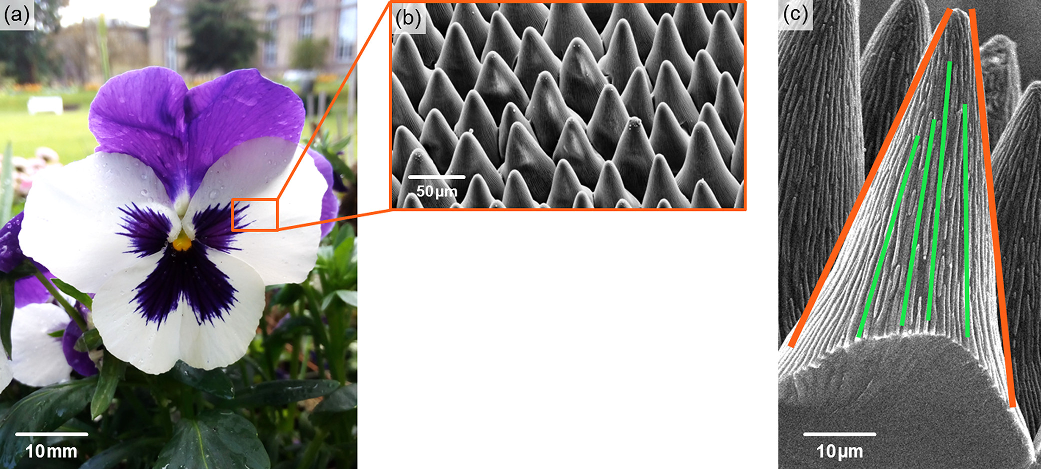}
    \caption{The petals of viola flowers (a) feature cone-shaped structures adorned with nanowrinkles (b,c). The cones and nanowrinkles collectively play a role in the augmentation of light harvesting. These figures were reproduced from ref.~\cite{Schmager2017}, with permission from the American Chemical Society.}
    \label{fig:Viola}
\end{figure}

In the realm of plants, surfaces of some petals and leaves reveal a mesmerizing array of structures optimised to enhance light harvesting (figure~\ref{fig:Viola}). The interplay of antireflection and light trapping mechanisms unfolds through the subtle architecture of conical-shaped epidermal cells~\cite{Gorton1996, Gkikas2015,Hunig2016,Schmager2017}. As sunlight encounters these structures, a gradual increase in the effective refractive index occurs, giving rise to antireflection akin to the cones found on the surfaces of moth eyes and cicada wings. In some plants, additional nanowrinkles were demonstrated to reduce light reflection~\cite{Hunig2016,Schmager2017}. In addition to antireflection, light redirection extends the path length within plant integuments, contributing to light trapping. The epidermal cells of flowers were reported to function as lenses and conducting incident light into the integuments comprising pigments~\cite{Gorton1996,Bone1985,Wilts2018}. It was also shown that the cone shape of their petals varies, reflecting the plant's strategic adaptation to either scatter or absorb incident ultraviolet waves, with shorter cones in the former, and taller ones in the latter, respectively~\cite{Schulte2019}. This adaptive variability plays a pivotal role in enhancing light capture for crucial processes like photosynthesis and contributes to the vivid coloration of these botanical wonders, especially in environments with limited light availability.

Venturing into the microscopic world, diatoms, unicellular algae encased in intricate silica frustules (namely, the hard porous structures of diatoms), offer a different yet equally captivating story of solar energy harvesting~\cite{Chen2015}. The case of \textit{Coscinodiscus} sp. stands out with its frustule comprising three layers – termed cribellum, cribrum, and the internal plate – each composed of thin slabs housing hexagonal arrays of disk holes. The size and spacing of these holes vary from layer to layer, forming a hierarchical structure that has been finely tuned for optimal light trapping and photosynthesis.

\begin{figure}
    \centering
    \includegraphics[width=135mm ]{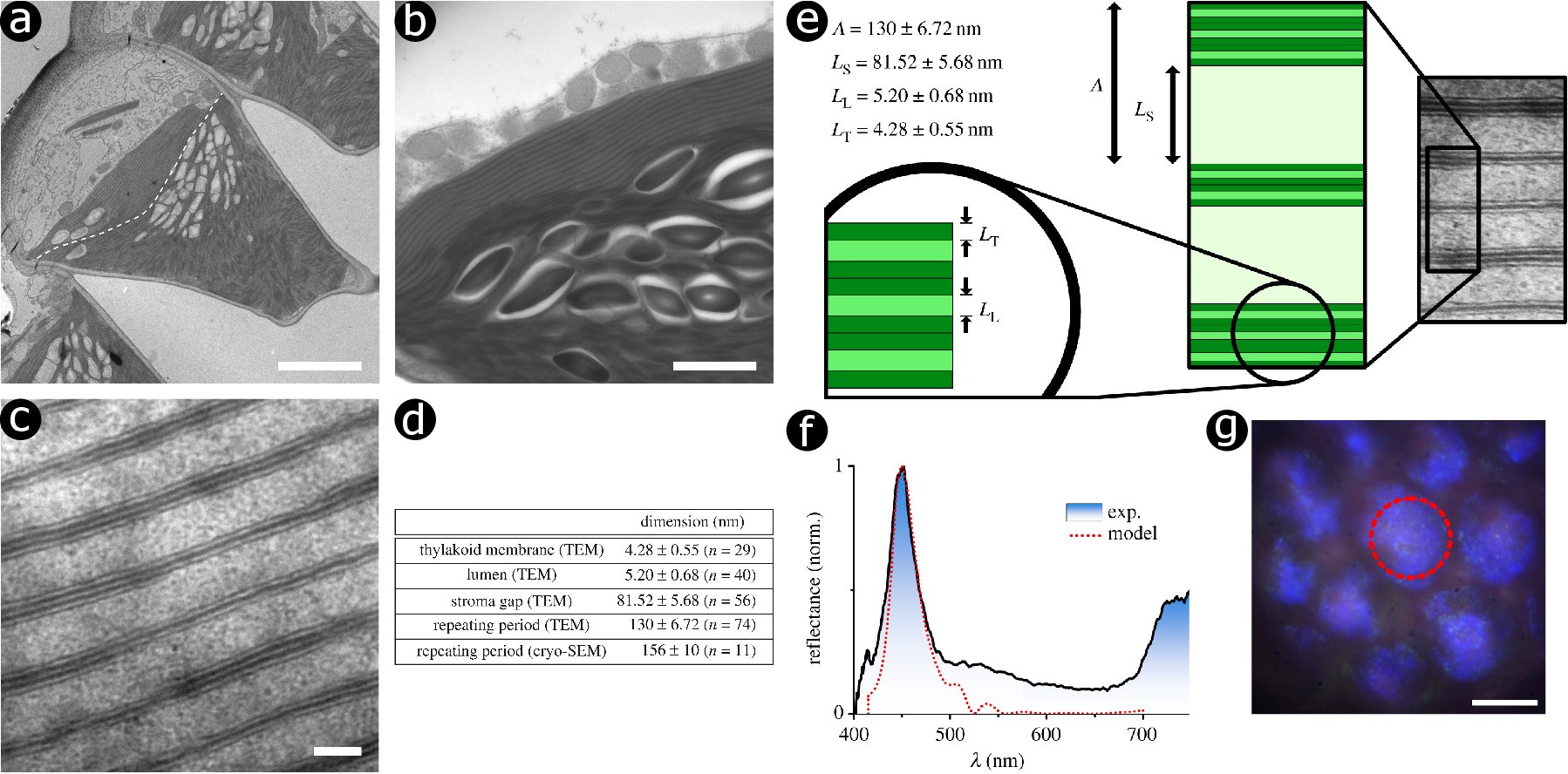}
    \caption{Bizonoplasts (a) found in the epidermis of the leaves of \textit{S. erythropus} contains very organised periodic multilayers (b,c) and conventional irregular thylakoid tissues, separated by a dashed line in (a). These SEM observations allow to measure the dimensions of these structures (d). The period $\Lambda$ of the multilayer comprises thylakoid membranes of thickness $L_\textrm{T}$, lumen layers of thickness $L_\textrm{L}$, and strommal layers of thickness $L_\textrm{S}$. Incident light reflects on this multilayer in the blue, as evidenced by spectrophotometry and simulations (f) as well as visualised through light microscopy (g). Scale bars: 5~$\mu$m (a), 0.5~$\mu$m (b), 100~nm (c), and 10~$\mu$m (g). These figures were reproduced from ref.~\cite{Masters2018}, with permission from the Royal Society.}
    \label{fig:selaginella}
\end{figure}

The blue iridescent epidermal chloroplasts occurring in some plant leaves, such as those in shade-dwelling \textit{Begonia} spp. and \textit{Selaginella erythropus}, display intricate multilayers that significantly enhance light absorption~\cite{Jacobs2016,Masters2018,Castillo2021,Wardley2022}. Chloroplasts, crucial plant organelles facilitating photosynthesis by absorbing incident light via chlorophyll, play a pivotal role in converting light energy into biochemical energy as adenosine triphosphate (ATP) and nicotinamide adenine dinucleotide phosphate (NADPH). Particularly, the initial light-dependent phase of photosynthesis takes place within the absorbing thylakoid tissues of chloroplasts. Two distinct types of chloroplasts are of interest with respect to photonics and enhanced light absorption: iridoplasts and bizonoplasts~\cite{Gould1996}. Iridoplasts, exclusive to some plants such as the leaves of \textit{Begonia}, possess photonic structures consisting in periodic multilayers of thylakoid tissues. Conversely, bizonoplasts have been identified in a single plant species, \textit{S. erythropus} (figure~\ref{fig:selaginella})~\cite{Sheue2007,Sheue2015}. They look like a mix of conventional thylakoid tissue present in typical irregular chloroplasts across many plants, and very organised iridoplasts. The unique ordered photonic structures of iridoplasts and bizonoplasts result in enhanced light absorption in the green part of the electromagnetic spectrum due to slow-light effect occurring at the red edge of the photonic bandgap of the multilayered structures. This increased absorption aligns with the incident light environment of these canopy-adapted plants. It leads to an enhanced quantum yield in low-light conditions, bolstering photosynthesis when compared to normal chloroplasts~\cite{Jacobs2016,Masters2018,Castillo2021,Wardley2022}.

\begin{figure}
    \centering
    \includegraphics[width=135mm ]{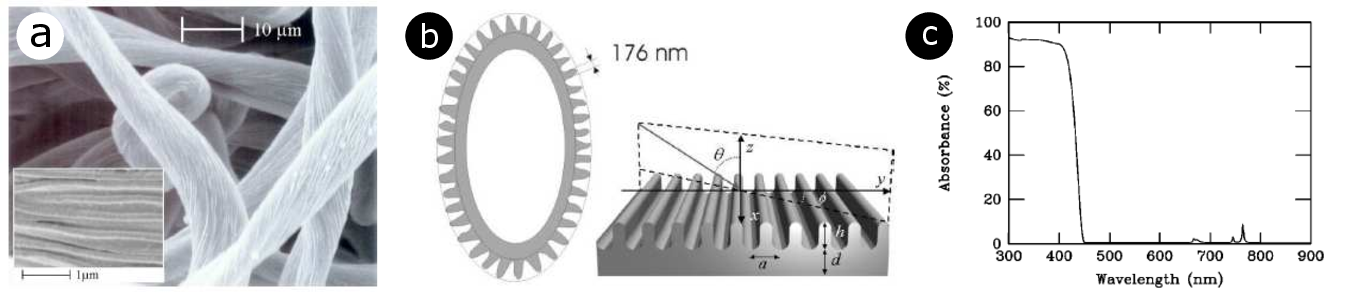}
    \caption{Protection against UV is a crucial attribute for edelweiss flowers (\textit{L. nivale}) as their silver-white bracts are exposed to Sun radiation. These bracts are enveloped by a downy layer consisting of an intersecting pattern of transparent filaments (a) with diameters of ca. 10~$\mu$m, as observed by SEM. These filaments display some corrugation (a, inset) with a period of ca. 180~nm. The total reflectance of a bract at normal incidence remains consistently high throughout the visible range but sharply diminishes to zero in the near UV - solid black line represents total reflectance, while the dashed red line represents the specular reflectance. The curvature of the hollow filament can be neglected in the models (b). The simulated absorbance spectrum through the corrugated-slab model for transverse-electric light polarisation exhibits a corresponding high intensity in the 300-450~nm range (c). These figures were reproduced from ref.~\cite{Vigneron2005}, with permission from the American Physical Society.}
    \label{fig:edelweiss}
\end{figure}

In the ethereal heights of the Alps, the edelweiss flower, \textit{Leontopodium nivale}, unveils a captivating defense mechanism against intense UV radiation. Located within the wooly cover layer of its bracts (namely, the downy-white “petals” of the edelweiss that actually are specialised leaves), an intersecting pattern of transparent filaments, displaying some faint iridescence can be observed by microscopy (figure~\ref{fig:edelweiss}). These filaments exhibit variations in diameter and morphology of their cross-sections. They are hollow with parallel corrugation running along the main axis of the filaments with a period of ca. 180~nm. The spectral reflectance is rather low from 300 to 400~nm and abruptly increases around 400~nm to form a plateau at ca. 65\%. This optical behaviour corresponds to some strong absorption in the near UV range. This intricate filamentary architecture, akin to a two-dimensional corrugated dielectric slab, emerges as a UV-selective waveguide coupling device. Fano resonances within these filaments facilitate the transfer of incident ultraviolet waves. The filaments, characterised by a broadband angular response, act as conduits for UV photon energy, efficiently dissipating it along the hollow guides, as the filament materials absorb UV. This ingenious strategy protects the delicate cellular tissue beneath.

\section{Infrared absorbers inspired by natural photonic structures}

The underlying mechanisms of various natural structures have been elucidated thus far, as presented in the previous section. If their potential for energy harvesting remains largely untapped, several devices have been suggested through a bioinspiration approach~\cite{Biro2011,Zhou2017,Mouchet2021}. In addition to their eco-friendly and sustainable materials composition and fabrication, these natural structures present other benefits with respect to current alternative such as their thinness and lightness. They can clearly improve the energy yield of PV cells and solar panels~\cite{Min2008,Han2015,Siddique2017}, passive radiative cooling~\cite{Didari2018,Zhang2020,Yang2021,Wang2023,Xu2024}, photocatalysis~\cite{Liu2013,Zhou2013,Yan2016}, or even the efficiency of electromagnetic camouflage, and the capture of stray light in telescopes.

\subsection{Bioinspired antireflective coatings: from moth-eye and cicada-wing templates to functional applications}

As described in section~\ref{winglepidopterans}, insects employ antireflective features as crucial characteristics in their crypsis strategy. The nipple arrays found on some of their eye and wing surfaces have been replicated to create bioinspired antireflective coatings applicable across various uses, including solar panels, antiglare glasses, screens, light-sensitive detectors, telescopes, thermochromic smart windows and camera lenses~\cite{Zhang2006,Linn2007,Min2008,Sun2008,Xie2008,Han2016,Tan2017,Zada2017,Liu2020,Mouchet2021,Novikova2024}.

\begin{figure}
    \centering
    \includegraphics[width=80mm ]{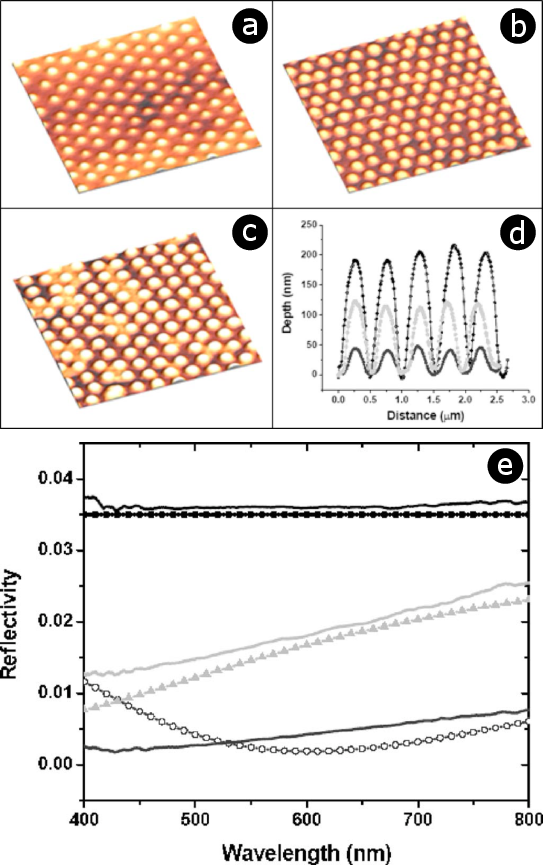}
    \caption{Nipple array structures observed on moth eye were replicated using 360-nm size silica particles, observed in (a-c) by atomic force microscopy (AFM) without etching (a) as well as with 20-s (b) and  45-s (c) RIE etching times. This etching process gave rise to diffent profiles (d). This bioinspired structuring gives rise to significant reduction in light reflection at normal incidence (e). Solid and dotted curves are respectively experimental and simulated reflectance spectra with a flat unstructured poly(ethoxylated trimethylolpropane triacrylate) (PETPTA) surface (black curves), 110-nm size hemispherical caps (light grey curves) fabricated with 20-s RIE etching, and 180-nm size hemispherical caps (dark grey curves) fabricated with 45-s RIE etching. These figures were reproduced from ref.~\cite{Linn2007} with permission from AIP Publishing.}
    \label{fig:Bottomtop}
\end{figure}

Bottom-up nanofabrication approaches such as self-assembled spherical nanoparticles in non-close-packed hexagonal arrays have been used to mimic moth-eye structures (figure~\ref{fig:Bottomtop})~\cite{Linn2007}. Simulations indicated that non-close-packed structures result in lower reflectance at wavelengths longer than the nipple interdistance, affirming the suitability of the natural moth-eye design for highly efficient antireflective devices. Spin-coating deposition of colloidal suspensions of silica particles (360~nm in diameter) and their shear alignment allowed to fabricate a template that served as a mould for casting some polydimethylsiloxane (PDMS). This PDMS mould was subsequently pressed onto a ethoxylated trimethylolpropane triacrylate (ETPTA)~\cite{Linn2007} or a perfluoroacrylate polymer~\cite{Sun2008a} layer laying on glass substrate (figure~\ref{fig:Bottomtop}a-d). These monomer films were then polymerised with a pulsed UV light curing system. Such biomimetic structures demonstrated outstanding low reflectance of less than 0.5\% across the visible spectrum (figure~\ref{fig:Bottomtop}e)~\cite{Linn2007,Sun2008a}. Similarly, such spin-coated monolayer silica colloids were utilised as a mask in a reactive ion etching (RIE) process of silicon wafers with SF$_6$~\cite{Sun2008} and of gallium antimonide (GaSb) substrates with Cl$_2$~\cite{Min2008}, reducing reflectance to less than 5\% in the visible-near-IR range (with respect to ca. 40\% for unstructured wafers)~\cite{Sun2008,Min2008}. Such biomimetic structures appeared easy to fabricate on solar and TPV cells.

\begin{figure}
    \centering
    \includegraphics[width=135mm ]{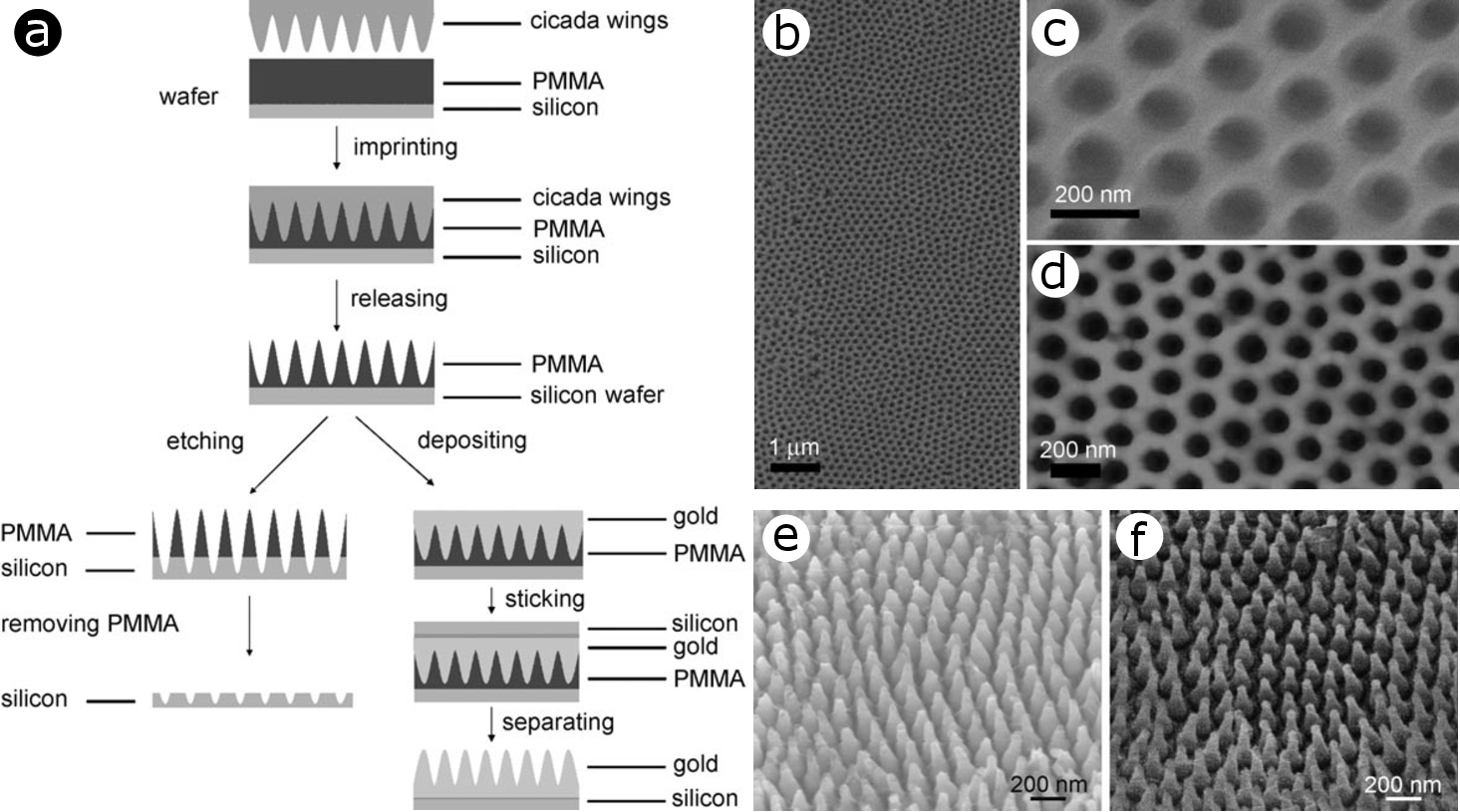}
    \caption{Nipple array structures were replicated in gold from cicada wings by NIL (a)~\cite{Zhang2006}. A negative replica was also fabricated in silicon (a). The process involved using cicada wings as natural stamp, giving rise to a PMMA negative replica observed by SEM (b,c) and AFM (d). Depositing gold on this PMMA mould allowed the fabrication of a perfect replica of the gold hexagonal nanopilars (e) with respect to the natural-wing structure (f). These figures were reproduced from ref.~\cite{Zhang2006} with permission from John Wiley and Sons.}
    \label{fig:topdown}
\end{figure}

Similarly, top-down synthesis techniques such as nanoimprint lithography (NIL) were employed to develop efficient and cost-effective antireflective coatings inspired from nature. For instance, cicada wings were directly utilised as natural stamps, leveraging the wing chitinous material with commendable thermomechanical characteristics (figure~\ref{fig:topdown}a)~\cite{Zhang2006,Xie2008}. This material can indeed be heated up to 200°C without any damage. After heating, a poly(methyl methacrylate) (PMMA) film may be pressed onto the biological template (figure~\ref{fig:topdown}b-d)~\cite{Zhang2006}. The negative replica was transferred to a silicon substrate using the array of nanowells in PMMA as a mask for RIE. Upon removal of PMMA, the resulting patterned surface demonstrated antireflective properties, evident from its dark visual appearance~\cite{Zhang2006}. If the structured PMMA film was employed as a mould for gold thermodeposition instead of a mask for RIE, a perfect replica of gold hexagonal nanopilars was produced (figure~\ref{fig:topdown}e,f)~\cite{Zhang2006}. Alternatively, a PMMA replica was fabricated through a modified procedure~\cite{Xie2008}: a first negative replica was obtained from the thermodeposition of gold onto the natural photonic structures of the wings. It was used as a mould to cast PMMA, which was subsequently peeled off~\cite{Xie2008}. The antireflective capability of the replicated PMMA film was notable. The PMMA film's reflectance was found to be decreased from approximately 6\% to about 2\% across the visible-near-IR range due to the unique nipple array~\cite{Xie2008}.

\subsection{Advances in bioinspired solar light harvesting: beyond petals, leaves and butterfly wings}

\begin{figure}
    \centering
    \includegraphics[width=135mm ]{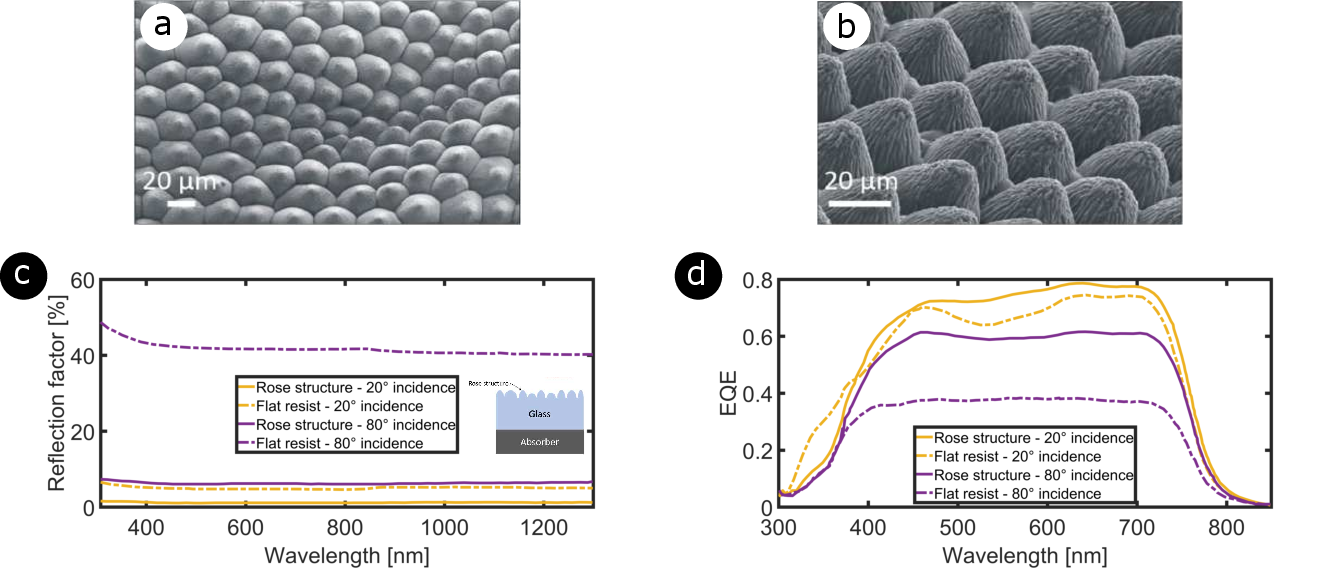}
    \caption{The microstructures occurring on the integument of rose petals were replicated in e.g., PMMA for application onto solar cells (a,b). They are observed here by SEM. The reflection factors from a glass substrate, featuring a black absorber on the rear and the replicated rose structure on the front side with light incident at angles of 20° and 80° (solid lines) are significantly lower than flat unstructured references (dash-dotted lines) (c). Consequently, the corresponding external quantum efficiencies (EQE) of the structured thin-film organic solar cell are higher than ones of the flat references (d). These figures were reproduced from ref.~\cite{Fritz2020b} (a,b), License CC-BY-4.0. Data for (c,d) were sourced from ref.~\cite{Hunig2016}.}
    \label{fig:rosa}
\end{figure}

The photonic structures found in the petals and leaves occurring in the integuments of certain plants have informed the development of improved bioinspired light-absorbing structures (figure~\ref{fig:rosa}a,b)~\cite{Liu2013,Zhou2013,Hunig2016,Schmager2017,Fritz2020a,Fritz2020b}. From enhancing the efficiency of organic solar cells to improving photocatalysis and contributing to artificial photosynthesis systems, these bioinspired designs continue to pave the way for sustainable and innovative energy solutions.

For instance, the microstructures of the epidermal cells of some rose species inspired a polymer thin film replica that was integrated in a solar cell~\cite{Hunig2016}. The biomimetic coating demonstrated a significant reduction in reflectance over the entire spectral range, particularly at grazing incidence, with remarkable 13\% and 44\% increase in the solar cell's short-circuit current at normal and grazing incidence, respectively (figure~\ref{fig:rosa}c,d). These properties are of high interest for solar cell efficiency with respect to the Sun movement throughout the day. The dome profile of the micropapillae on the petal surface was demonstrated to play the role of microlenses, lengthening the optical path of light rays within the plant integuments. This dual functionality of efficient antireflection and light trapping is crucial for enhancing the performance of thin-film organic solar cells, addressing issues of low optical absorption and spectral drops due to Fabry-Pérot interferences.

While the petal of roses offer a compelling template for biomimetic light harvesting, the sophistication of plant leaves provides an even more complex blueprint. The thin and soft leaves of \textit{Vallisneria} spp., aquatic grass plants, also known as eelgrass are an very informative study case for bioinspiration~\cite{Liu2013}. The hierarchical architecture of these leaves includes lens-like epidermal cells, a so-called palisade parenchyma functioning as optical waveguides, and a spongy disordered layer with intertwined veins giving rise to optical multiscattering and extending the optical path length. Utilising a sol-gel method, a silica and titania mimic of these leaves was templated for photocatalitic application~\cite{Liu2013}. The resulting Ti-Si catalyst exhibited a threefold higher rate constant for the degradation reaction of methylene blue exposed to UV compared to a commercial TiO$_2$ catalyst. The macroporosity and enhanced light scattering properties of \textit{Vallisneria} leaf structure made it an ideal template for photocatalysis, showcasing the potential for biomimicry in advancing solar-driven environmental applications.

\begin{figure}
    \centering
    \includegraphics[width=135mm ]{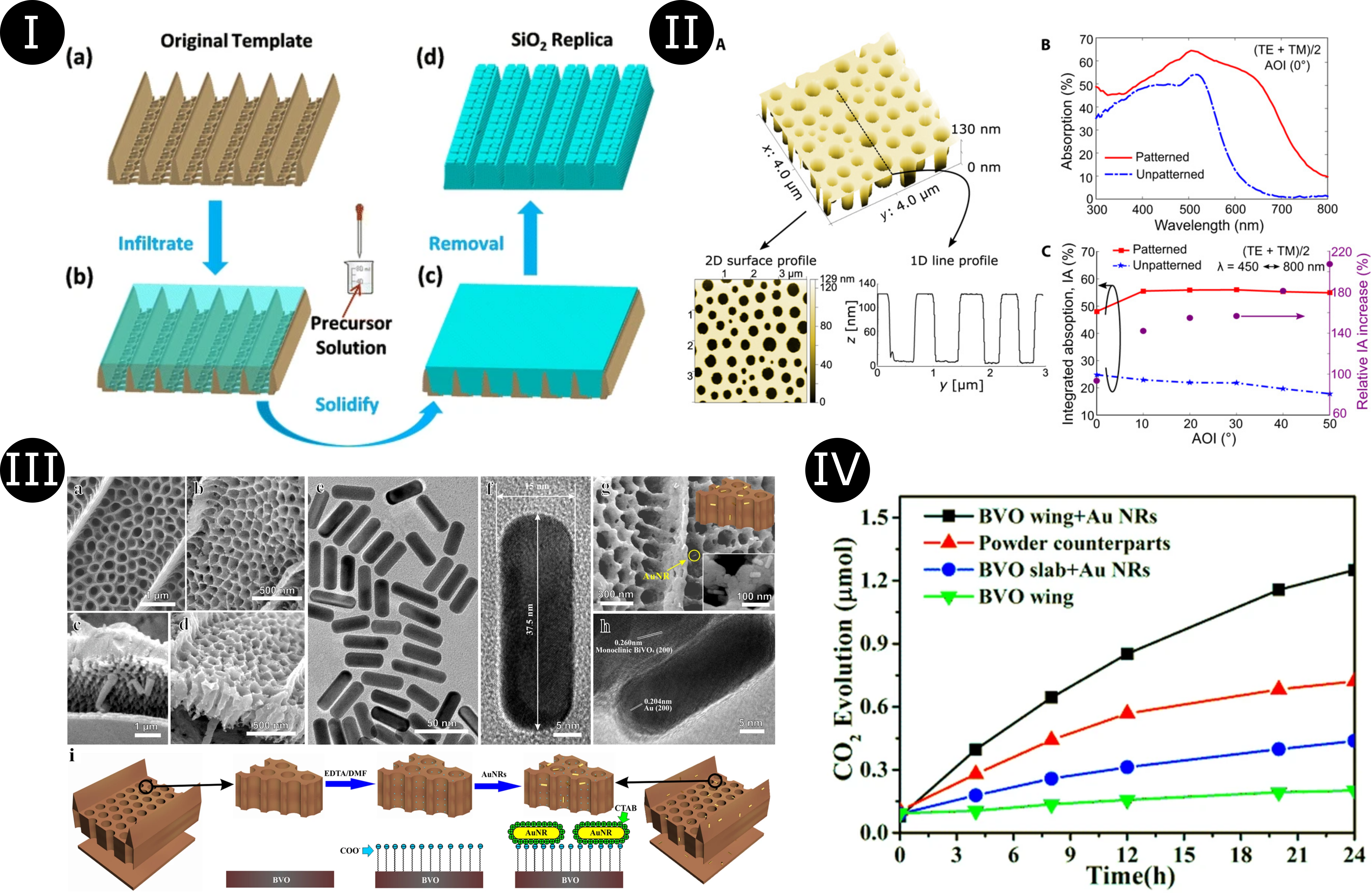}
    \caption{ (I) SiO$_2$ negative replica of the black wing scales of \textit{Trogonoptera brookiana} butterfly (a) was synthesised through a sol-gel method~\cite{Han2015}. The scales were infiltrated with a precursor solution (b) that was heated for solidification (c). The scales were removed by etching (d), leaving a negative replica. (II) A sol-gel method was also used to fabricate a bismuth vanadate (BVO) replica (b,d) of the wing scales of \textit{Papilio nireus} (a,c)~\cite{Yan2016}. Gold nanoantennas (abbreviated "AuNR") (e,f) were loaded into the BVO wing scales (g,h). Insets in (g) are a simplified sketch (top) and a higher-magnification electron micrograph (bottom). The fabrication process from the butterfly wing scale to the Au NR-loaded BVO hybrid photonic-plasmonic structure is summarised in (i). (III) This hybrid structure gave rise to the best photocatalytic activity~\cite{Yan2016}, as demonstrated by the CO$_2$ evolution of isopropyl alcohol (IPA) degradation as a function of illumination time for the Au NR-loaded BVO hybrid photonic-plasmonic structure ("BVO wing+Au NRs"), unstructured BVO powder ("Powder counterparts"), unstructured BVO slab with Au NRs ("BVO slab+Au NRs"), and bio-templated BVO photonic structure ("BVO wing"). (IV) A nanostructured hydrogenated amorphous silicon (a-Si:H) film inspired by the wing scales of \textit{Pachliopta aristolochiae} (a) gave rise to enhanced light harvesting at normal incidence (b) and with no-zero incidence angles (c)~\cite{Siddique2017}. These figures were reproduced from ref.~\cite{Han2015} (I), ref.~\cite{Yan2016} (II,III), and ref.~\cite{Siddique2017} (IV), Licenses CC-BY-4.0, CC-BY, and CC-BY-NC, respectively.}
    \label{fig:butterfly_mimic}
\end{figure}

As one could have expected, butterfly wings have naturally inspired the design of visible and IR light absorbers such as an SiO$_2$ negative replica of the black wing scales of \textit{Trogonoptera brookiana} butterfly that exhibits enhanced light trapping properties (figure~\ref{fig:butterfly_mimic}I)~\cite{Han2015}. The solar energy loss of this replica, namely, the integrated solar light intensity reflected by the replica in the 400-900~nm range, was found to be 22.6\% of the one of an SiO$_2$ flat surface. Similarly, a hybrid photonic-plasmonic structure was fabricated by bio-templating the black forewings of \textit{Troides helena}~\cite{Tian2015}. Silver spherical nanoparticles with various diameters (10~nm, 20~nm, 40~nm, 60~nm, and 80~nm) were deposited on the wings before the chitin structure was carbonised. An enhanced absorption was measured and simulated in the near- and mid-IR ranges. It results from the plasmon resonance of the silver nanoparticles and the coherent coupling among adjacent nanoparticles within the photonic architecture of \textit{T. helena}. Such a hybrid photonic-plasmonic architecture was designed for photocataltic applications while taking inspiration from \textit{P. nireus} (figure~\ref{fig:butterfly_mimic}II)~\cite{Yan2016}. It consisted in gold nanoantennas located onto a bismuth vanadate (BVO) photocatalytic unit with the architecture of \textit{P. nireus} black wings. This architecture was fabricated through a sol-gel method. Both experiments and simulations demonstrated the enhanced photocatalytic activity arising from the 25\%-increased light harvesting within the 700~nm-1200~nm range and the 3.5-fold enhancement of the electric-field intensity of localised surface plasmons (figure~\ref{fig:butterfly_mimic}III)~\cite{Yan2016}. Whereas these three artificial structures were fabricated by bottom-up methods, a nanostructured absorber film with disordered holes was synthesised as a mimic of the disordered black wing scales of \textit{P. aristolochiae} butterfly (figure~\ref{fig:butterfly_mimic}IV)~\cite{Siddique2017}. Using phase separation of a two-polymer mixture, a hydrogenated amorphous silicon (a-Si:H) film was patterned for PV applications. The structure exhibited a relative integrated absorption over the range 450~nm-800~nm of 93\% and 207\%, with a 0°- and a 50°-incidence angle with respect to the normal to the film surface, respectively~\cite{Siddique2017}.

\section{Conclusions}

Photonic structures occurring in the integuments of living organisms such as arthropods, birds, and plants are very sophisticated optical devices that give rise to various striking optical effects, including UV, visible, and IR radiation management and absorption enhancement. They occur in biological tissues encompassing butterfly wings, beetle elytra, bird feathers, spider cuticle, viper skin, as well as plant leaves and petals. These phenomena are often crucial for the survival of animal and plant species. This review article also showcased the promising potential of bioinspiration in the field of energy capture and conversion. Exploiting light trapping, impedance matching or antireflection observed in natural structures is indeed highly interesting in view of developing bioinspired energy-efficient applications such as PV and TPV cells, TEG, artificial photosynthesis, and photocatalysis. With the optimisation of the efficiency of such applications, these advances inspire future research and innovation in the field of bioinspired energy solutions. Ultimately, this research paves the way for a more sustainable and environmentally conscious future by harnessing the beauty of nature's designs to meet humankind's energy needs.

\section*{Acknowledgements}

The author thanks Prof. Olivier Deparis from the University of Namur, Belgium for fruitful discussion and reading the manuscript. The author was supported by a BEWARE Fellowship (Convention n°2110034) of the Walloon Region (COFUND Marie Skłodowska-Curie Actions of the European Union \#847587).

\section*{Conflicts of interest}

The author declares no conflict of interest.

\bibliographystyle{unsrt} 
\bibliography{bibliography.bib}

\end{document}